\documentclass[a4paper,fleqn]{cas-dc}

\usepackage[authoryear,longnamesfirst]{natbib}

\usepackage[utf8]{inputenc}
\usepackage{textcomp}

\usepackage{graphicx}
\usepackage{caption}
\usepackage{subcaption}
\usepackage{savesym}
\usepackage{amsmath}
\savesymbol{Bbbk}
\usepackage{amssymb}
\restoresymbol{SYM}{Bbbk}
\usepackage[version=4]{mhchem}
\usepackage{siunitx}
\usepackage{longtable,tabularx}
\usepackage{bm}
\usepackage{color}
\usepackage{dsfont}
\usepackage{multicol}

\usepackage[english]{babel}
\addto\captionsenglish{}
\usepackage{theoremref}
\usepackage{hyperref}
\usepackage{float}
\usepackage{url}
\usepackage{booktabs}

\newtheorem{thm}{Theorem}

\newtheorem{prob}{Problem}

\newtheorem{prf}{Proof}

\def\tsc#1{\csdef{#1}{\textsc{\lowercase{#1}}\xspace}}
\tsc{WGM}
\tsc{QE}

\begin{document}
\let\WriteBookmarks\relax
\def\floatpagepagefraction{1}
\def\textpagefraction{.001}

\shorttitle{Modeling, Simulation and Maneuvering Control of a Generic Submarine}

\shortauthors{Modeling, Simulation and Maneuvering Control of a Generic Submarine}  

\title [mode = title]{Modeling, Simulation and Maneuvering Control of a Generic Submarine}  

\tnotemark[1] 

\tnotetext[1]{This research was supported by the Office of Naval Research (grants N000142212634 and N000142112091).} 

%

\author[1]{Gage MacLin}






\affiliation[1]{organization={University of Iowa},
            addressline={103 S. Capitol Street}, 
            city={Iowa City},
            postcode={52242}, 
            state={Iowa},
            country={United States}}

\author[1]{Maxwell Hammond}








\author[1]{Venanzio Cichella}





\author[1]{Juan E. Martin}









\begin{abstract}
This work introduces two multi-level control strategies to address the problem of guidance and control of underwater vehicles. An outer-loop path-following algorithm and an outer-loop trajectory tracking algorithm are presented. Both outer-loop algorithms provide reference commands that enable the generic submarine to adhere to a three-dimensional path, and both use an inner-loop adaptive controller to determine the required actuation commands. Further, a reduced order model of a generic submarine is presented. Computational fluid dynamics (CFD) results are used to create and validate a model that includes depth dependence and the effect of waves on the craft. 
Results from the reduced order model for each control strategy are compared.
\end{abstract}



\begin{keywords}
Path following \sep Trajectory tracking \sep Underwater vehicle control \sep Joubert BB2
\end{keywords}

\maketitle

\section{Introduction}

Reduced order models (ROM) are a suitable option for the development of control algorithms and path planning for submarines. ROMs represent a trade-off between accuracy and execution cost but, if able to correctly reproduce the dynamic response of the vehicle, are an excellent tool to study maneuvering under external disturbances, such as waves \citep{fang2006wave}, or for collision avoidance \citep{jin2020dynamic}, in particular if multiple simulations are required for optimization. The ROM used in this work has been developed explicitly to consider wave and depth effects and to support the development of advanced controllers \citep{rober2021three}.

Modeling operation near the surface, in restricted waters, and at low speeds is a particularly challenging problem as craft controllability becomes compromised by the reduced authority of its control surfaces or more demanding requirements are present as danger
of collision or surfacing increase. The development of controllers and control strategies is greatly aided by dynamic models of the craft that are fast, yet accurate. While experimental or high-fidelity computations can also be used for testing, they are considerably more expensive and time consuming. Early development of control strategies and algorithms, and subsequent initial tuning are typically performed using ROM, leaving experiments and CFD for fine tuning or to study \textcolor{red}{and further validate} controller performance under more complex effects not captured by simpler ROM approaches. 

The ROM solves the rigid body equations of motion of the vehicle under external forces and moments. The external forces in a submarine include the hydrostatic and hydrodynamic forces. The hydrodynamic forces are a result of the state of motion of the vessel, in the form of virtual mass, pressure drag, and skin friction; the objective of the ROM is to relate those forces to the kinematics of the craft. Accurate prediction of the hydrodynamic loads requires information under wide motion conditions and the calibration of a large number of coefficients. The most frequently used form of hydrodynamic model (HDM) for cruciform stern plane configurations was proposed by \cite{gertler1967standard}. In this HDM only forces and moments due to certain motions are considered and the history of motions is neglected. The coefficients or derivatives required by the model are evaluated from experiments, semi-empirical approaches, or computational methods.

Experimental methods are the most conventional technique to obtain the coefficients by performing captive model tests with scaled models. Towing with or without Planar Motion Mechanisms (PMM) and Rotating Arms (RA) are used to isolate the desired coefficients \citep{feldman1995method}. For submarines, wind tunnel tests are also possible for deep conditions. The validation of HDM can be done experimentally by performing free running controlled maneuvers in model scale in a wave basin \citep{overpelt2015free}.  Experimental techniques provide reliable data but require expensive facilities and construction of a model. In addition, experiments are limited to model scale and require extrapolation of results  to full scale.

Computational techniques can also be used for coefficient evaluation. Potential flow solvers are cost-effective tools to estimate pressure effects and virtual mass. However, this methodology is not accurate for separated flows, providing poor estimates at large angles of attack \citep{evans2004dynamics}. CFD can also be used for this purpose, however the simulation cost may limit its use to relatively simple geometries or require the use of very coarse grids. In the past, unsteady tests such as PMM might have been replaced by static calculations to reduce the computational requirements. Current capabilities have allowed the simulation of more challenging conditions, and full maneuvers are routinely completed using moving control surfaces \citep{carrica2021cfd}. The success in predicting maneuvers with CFD lends credibility to its use to obtain the coefficients of ROM, but development of accurate ROM simulation models is a significant challenge as the operational conditions deviate from those used to obtain the coefficients.

A reduced order model  for the generic submarine Joubert BB2 is presented in this paper.  The model originally proposed by \cite{gertler1967standard} has been updated to include an X-shape for the stern planes and a simplified formulation of some of its terms. More importantly, surface effects and added mass effects due to wave action have been included in the model, allowing the simulation of maneuvers near the surface. Extensive computational fluid dynamics  tests were conducted using REX, a CFD solver developed at The University of Iowa, to generate the model coefficients. The resulting model is implemented in the commercial software MATLAB SIMULINK\texttrademark   with the purpose of generating an open model for a variety of applications. The resulting hydrodynamic model (HDM) has been satisfactorily compared to available experimental and numerical data for Joubert BB2 \citep{carrica2019,carrica2021cfd} and used to support the development of novel controllers. The model provides a quick platform for evaluation of control strategies, planning of maneuvers, etc. 
As an example of the capabilities of the ROM and the controllers, a \textcolor{red}{3}DoF path-following maneuver and a \textcolor{red}{3}DoF trajectory tracking maneuver simulated using the ROM are presented. The SIMULINK model along with inputs for the validation case presented in this paper and others can be found at our github web page \footnote{\url{https://github.com/caslabuiowa/IowaBB2model}}.

In order for underwater vehicles to conduct specific maneuvers, it is imperative that they are capable of following desired spatial trajectories. One of the most common methods for trajectory adherence are trajectory tracking algorithms (e.g., \cite{guerrero2019trajectory,henninger2019trajectory,gong2021lyapunov}). These algorithms calculate the required actuation commands for the vehicle to adhere to a specified trajectory, and to accomplish this trajectory in a specified time. An alternative are path-following algorithms (e.g, \cite{abdurahman2017switching,peng2017output,paliotta2018trajectory,encarnacao20003d,lapierre2003nonlinear}), which are similar to trajectory tracking algorithms in that they also calculate the required actuation commands to adhere to a specified trajectory, or path. However, path-following algorithms are not constrained temporally. Trajectory tracking is directly parameterized by time, which introduces a temporal constraint. This trait proves valuable for time-critical missions. Path following differs from trajectory tracking in that it isn't directly parameterized by time, instead being parameterized by a virtual time. This provides an additional degree of freedom to the system, which allows the vehicle to follow the path using many different velocities. This provides the user the opportunity to directly set the velocity. The path-following algorithm introduced in this work borrows from \cite{rober2021three,rober20223d}, however the path-following algorithm used here is reformulated such that depth and heading (yaw) are directly controlled, instead of directly controlling angular rates (pitch and yaw rates), \textcolor{red}{which changes the formulation of the path-following algorithm to accommodate for the augmentation of more general underwater vehicle autopilots.}

While there are many different outer-loop control strategies to generate desired reference signals, each utilize a vehicle autopilot which calculates the required vehicle dynamics to follow the desired reference. 
A common method for autopilot design is proportional-integral-derivative (PID) control. 
The inner-loop control strategy in this work utilizes $\mathcal{L}_1$ adaptive control, presented in \cite{hovakimyan2010}, to augment the reference signal to estimate unknown disturbances and improve the performance of the system. Thus, the first control architecture to be presented utilizes an inner-loop $\mathcal{L}_1$ adaptive controller to augment a PID autopilot, which receives reference commands from an outer-loop path-following algorithm. The second control architecture uses the same inner-loop controller, but uses a trajectory tracking algorithm as an outer-loop controller. Both of these architectures are used in simulation to maneuver the generic submarine model Joubert BB2 introduced in \cite{carrica2019}. \textcolor{red}{The novelties of this work include: (1) the formulation of path following and trajectory tracking algorithms that leverage existing submarine autopilots capable of processing depth, pitch, heading angle, and speed commands; (2) the development of an adaptive controller that enhances these autopilots, offering increased robustness and augmented performance in the presence of disturbances and diminished autopilot performance; (3) a direct case-by-case comparison between path-following and trajectory tracking algorithms for an underwater vehicle; and, (4) the presentation of a ROM including near surface effects and varying sea states that allows for rapid development and initial validation of control algorithms for underwater vehicles. This work, including the path-following, trajectory tracking, and adaptive control algorithms, as well as the ROM are available on our github.}

{This paper is organized as follows. In Section \ref{sec:methods}, we introduce the Joubert BB2 geometry and the hydrodynamic model.
In Section \ref{sec:controller}, we formulate the path-following and trajectory tracking problems, and introduce the underlying adaptive controller. Section \ref{sec:results} then demonstrates the controllers performance through a series of simulations using the HDM as well as CFD, which shows how each control strategy performs in various simulations. Next, Section \ref{sec:conclusion} summarizes the work presented in this paper, and outlines further work.}

\section{Modeling}\label{sec:methods}

\subsection{Joubert BB2}
Joubert BB2 \citep{joubert2006some,overpelt2015free} is used for this work. The main particulars at prototype scale are presented in Table~\ref{tab:BB2}. The model includes a X-shape arrangement of control stern planes and a set of sail planes. The original control of BB2 consisted of a  pair of PD controllers for direction, that combine control of depth and pitch in a vertical command $\delta_V$ and of lateral displacement and yaw in a horizontal command $\delta_H$. The control planes are then actuated as combination of the two commands:
\begin{equation}
\begin{split}
    \delta_1&= -\delta_V + \delta_H  \text{\hspace{1cm}(lower starboard)}\\
    \delta_2&= -\delta_V - \delta_H \text{\hspace{1.22cm}(upper starboard)}\\
    \delta_3&= \delta_V - \delta_H \text{\hspace{1.22cm}(upper port)}\\
    \delta_4&= \delta_V + \delta_H  \text{\hspace{1cm}(lower port)}\\
    \delta_5&= \delta_V   \text{\hspace{1.76cm} (sail)}\\
   \end{split}
    \label{CommandsVH}
\end{equation}
 
\begin{table}
    \centering
    \caption{Main particulars of Joubert BB2 (prototype scale)}
    \begin{tabular}{l l c}
\toprule
\!Length&$L_0$ (m)&70.2 \\
\!Beam&$B$ (m)&9.6\\
\!Depth to top of sail &$D_0$  (m)&16.2\\
\!Displacement&$\nabla$ (tons)&4440\\
\!Center of Gravity \\ \quad Long. (from nose)&$X_G$  (m)&32.31\\
\quad Vertical  (from shaft)&$Z_G$  (m)&0.0443\\
\!Gyration radii\\ \quad Roll&$r_x$  (m)&3.433\\
\quad Pitch &$r_y$  (m)&17.6\\
\quad Yaw &$r_z$  (m)&17.522\\
\bottomrule
    \end{tabular}
    \label{tab:BB2}
\end{table}

\subsection{Hydrodynamic Model}\label{ExpModel}
\begin{figure}
  \centering
    \includegraphics[width=.35\textwidth]{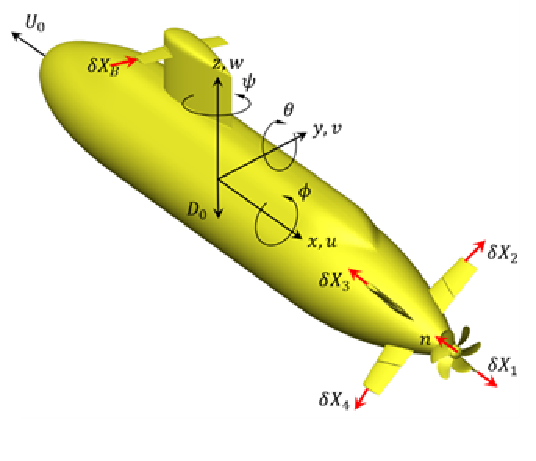}
    \caption{Definition of axes and variables used for modelling BB2.}
    \label{fig:axis_def}
\end{figure}

The ROM solves the six degrees of freedom (6DoF) equations of motion of the craft in a coordinate system local to the body: 
\begin{equation} \label{eq:sixdofgen} 
M \dot{s} = F-b=F_b+F_h-b,  
\end{equation} 
where $s= [u,v,w,p,q,r]^T$ is the generalized velocity vector, $M$ is the mass matrix, given as a function of the mass $m$, the center of gravity of the craft $(x_G,y_G,z_G)$ and the inertia tensor $I$ as: 
\begin{equation} \label{eq:M}
\footnotesize{
 	\!M\!=\!\begin{bmatrix} m&0 & 0 & 0 &m z_G & -my_G \\ 
 	                0&m & 0 & -mz_G &0 & mx_G \\ 
 	                0&0 & m & my_G &-mx_G & 0 \\ 
 	                0&-mz_G & my_G &I_{xx} &-I_{xy} & -I_{xz} \\ 
 	                mz_G&0 & -mx_G &-I_{xy} &I_{yy} & -I_{yz} \\ 
 	                -my_G&mx_G & 0 &-I_{xz} &-I_{yz} & I_{zz} \\ 
 	                \end{bmatrix}\!}
\end{equation}
Since the equations are solved in the ship system (shown in Fig.~\ref{fig:axis_def}), all extra diagonal terms are typically zero but are included in the general model to consider changes to the mass distribution through actuation of trim tanks for control. The inertia tensor diagonal terms are obtained from the gyration radii as $I_{ii} = m r_i^2$. Extra-diagonal terms are considered null. 
The coupling terms due to the use of a non-inertial reference system, $b$ is 
\begin{equation} \label{eq:b}
\small{
 b=	\begin{bmatrix} m[wq\!-\!vr\!-\!x_G (q^2\!+\!r^2 )\!+\!y_G (pq\!-\!\dot{r} )\!+\!z_G (pr\!+\!\dot{q} )]\\
 	                m[ur\!-\!wp\!-\!y_G (r^2\!+\!p^2 )\!+\!z_G (qr\!-\!\dot{p} )\!+\!x_G (qp\!+\!\dot{r} )]\\
 	                m[vp\!-\!uq\!-\!z_G (p^2\!+\!q^2 )\!+\!x_G (rp\!-\!\dot{q} )\!+\!y_G (rq\!+\!\dot{p} )]\\
 	                  (I_{zz}\!-\!I_{yy})qr\!+\!m[y_G (\dot{w}\!-\!uq\!+\!vp)\!-\!z_G (\dot{v}\!-\!wp\!+\!ur)]\\
 	                  (I_{xx}\!-\!I_{zz})rp\!+\!m[z_G (\dot{u}\!-\!vr\!+\!wq)\!-\!x_G (\dot{w}\!-\!uq\!+\!vp)]\\
 	                  (I_{yy}\!-\!I_{xx} )pq\!+\!m[x_G (\dot{v}\!-\!wp\!+\!ur)\!-\!y_G (\dot{u}\!-\!vr\!+\!wq)] \\
 	                  \end{bmatrix}}
\end{equation}
The external loads $F$ are split in hydrostatic $F_b$ and hydrodynamic forces $F_h$. Craft weight $W$ and buoyancy $B$ are included in $F_b$, considering the possibility of changes to these terms due to control mechanisms that change total mass of the craft (ballast tank), or its distribution (trim tanks):
\begin{equation} \label{eq:F_b}
\small{ 	F_b \!=\!
 	\begin{bmatrix}
 	-(W-B)  \sin{\theta}\\
 	  (W-B)  \cos{\theta}  \sin{\phi}\\
 	  (W-B)  \cos{\theta} \cos{\phi} \\
 	  (y_G W\!-\!y_B B)\cos{\theta}\cos{\phi}\!-\!(z_G W\!-\!z_B B)\cos{\theta} \sin{\phi}\\
 	  (z_G W\!-\!z_B B) \sin{\theta}\!-\!(x_G W\!-\!x_B B)\cos{\theta}\cos{\phi}\\
 	  (y_G W\!-\!y_B B) \sin{\theta}\!-\!(x_G W\!-\!x_B B)\cos{\theta}\sin{\phi}\\	                  \end{bmatrix}}
\end{equation}
The external load $F_h$
 parameterizes the loads  acting on the craft in terms of relevant kinematic variables such as craft speed, and angle of attack (AoA) of the craft and its control surfaces, as well as external parameters, including distance to the surface and sea state.   The general form of the six forces and moments $F_{h,i} $ is
\begin{equation} \label{eq:sixdof_fh} 
\begin{split}
F_{h,i} =\sum_{k=1}^6\sum_{j=1}^6 F'_{i,s_j s_k} s_j s_k + \sum_{j=1}^6 F'_{i,s_j} \dot{s}_j +\\ \sum_{l=1}^5 F'_{i,\delta_l} u^2 \delta_l^2+F_{i,prop}.
\end{split}
\end{equation} 
The hydrodynamic loads are grouped in four terms. The first group includes drag terms and added mass effects due to motions. The second group corresponds to added mass due to acceleration. Virtual mass and wave effects are considered by discretizing the boat geometry into sections to generate lumped coefficients for each section. The third group corresponds to the forces and moments generated by the effective deflection 
$\delta_l$ of each control surface (four stern planes in cruciform shape, plus the sail plane); and finally the propeller forces and moments  are also considered.

\begin{figure}
  \centering
    \includegraphics[width=.49\textwidth]{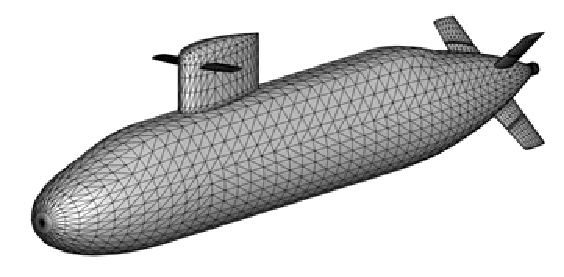}
    \caption{Triangulated surface for hydrostatic load calculation. The total number of elements $N_e$ is 7,148.}
    \label{fig:hydrostat}
\end{figure}

Wave-induced loads are incorporated to the model by integration over a coarse grid as shown in Fig.~\ref{fig:hydrostat}. While the wave pressure force is dynamic, it is herein considered independent from the state of the vehicle and thus only a function of time and space, allowing grouping with the hydrostatic force. For fully submerged, neutrally buoyant conditions, the weight is adjusted to exactly match the buoyancy obtained by integration as initial condition, and allowed to evolve if a ballast controller is used, and restoring moments are calculated as shown in Equation~\eqref{eq:F_b}. The additional forces and moments due to waves are calculated by integration of the pressure at the Gauss points of the grid shown in Fig.~\ref{fig:hydrostat} as 
\begin{equation} \label{eq:hydrostat_mom} 
\begin{split}
F_{w,i} &=-\frac{1}{3}\sum_{i=1}^{N_e}(p_{12,i}+p_{23,i}+p_{31,i})A_i;\\
M_{w,i}\! &=\!-\frac{1}{3}\sum_{i=1}^{N_e}(r_{12,i}p_{12,i}\!+\!r_{23,i}p_{23,i}\!+\!r_{31,i}p_{31,i})\!\times\! A_i;\\
\end{split}
\end{equation} 
with $A_i$ the area vector of each element and $r_{i}$ the coordinates of the Gauss point in the ship system. The pressure field is given analytically as that of a progressive regular wave in deep water

\begin{equation} \label{eq:wave} 
\begin{split}
p\! =\!-\rho g z + \rho g A_{wave} e^{k_{wave}z} \sin(k_{wave} x - \omega_{wave} t)
\end{split}
\end{equation} 
where $k_{wave}$ is the wave number, $\omega_{wave}$ is the wave frequency in rad/s, $A_{wave}$ is the amplitude and $z$ is the vertical distance to the calm water level. This method can be extended to include more complex wave fields.  

\subsection{Computational Fluid Dynamics}\label{CFD}

The coefficients in Eqs.~\eqref{eq:sixdof_fh} need to be modeled from existing numerical or experimental data. In the present work, high-fidelity computational fluid dynamics results, generated with REX is used. REX solutions for BB2 have been extensively validated against experimental data \citep{carrica2021cfd}.
 REX is a  hybrid RANS/LES solver based on the SST turbulence model \citep{menter1994two} with dynamic overset capabilities \citep{noack2009suggar++}. The free surface, included in the calculations used to develop the HDM, is modelled using single-phase level set \citep{carrica2007ship}. Further details on REX capabilities and numerical implementation can be found in \citep{li2020modeling}.

 A coarse multiblock mesh was used for coefficient calculation and for validation runs. The mesh allows the motion of discretized control surfaces, but does not include a discretized propeller, which is instead replaced by a body force. The coarse grid (4.5 M grid points) has been used in the past for validation purposes \citep{carrica2021vertical} and it is known to produce acceptable results. While a finer grid could be used to reduce errors in the coefficients estimates, it is expected that the main source of inaccuracies is in the ROM approach itself, and thus the additional numerical cost is not justifiable.

\section{Control}\label{sec:controller}
\begin{figure}
    \centering
     \includegraphics[width=0.49\textwidth]{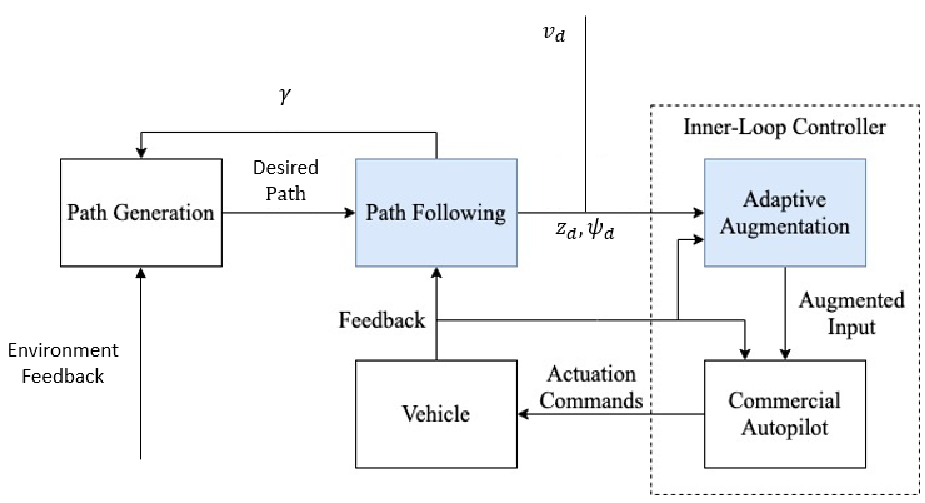}
    \caption{The overall control structure highlighting the path-following controller and the adaptive augmentation algorithm}
    \label{fig:pf_block_diagram}
\end{figure}

\begin{figure}
    \centering
     \includegraphics[width=0.49\textwidth]{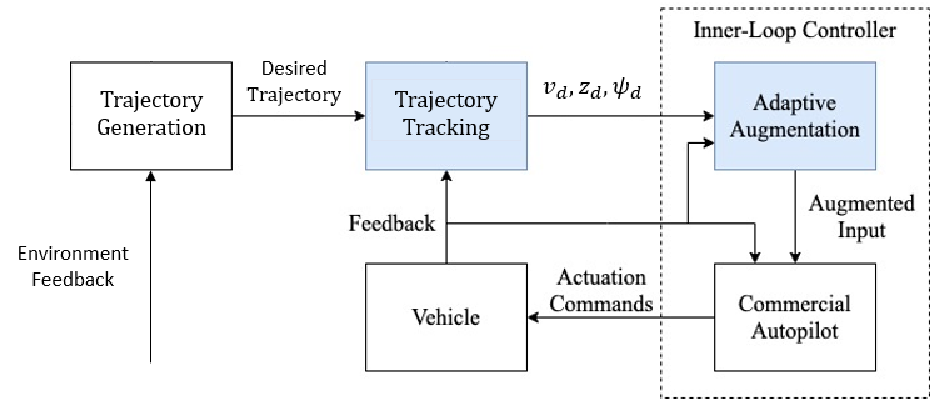}
    \caption{The overall control structure highlighting the trajectory tracking controller and the adaptive augmentation algorithm}
    \label{fig:tt_block_diagram}
\end{figure}

The Joubert BB2 reduced order model has been used as a basis for validation and testing of two multi-layer control architectures. The architectures are illustrated in Figures ~\ref{fig:pf_block_diagram} and \ref{fig:tt_block_diagram}. Both architectures require motion planning algorithms to generate a desired trajectory or path for the vehicle to follow. Additionally, an on-board autopilot able to track speed, depth, and yaw commands is necessary for both architectures. With this assumption, we can formulate and address the path-following and trajectory tracking problems in the 2D horizontal plane, while allowing the autopilot to directly control the desired depth. 

The path-following architecture, shown in Fig.~\ref{fig:pf_block_diagram}, operates as follows: $(i)$ the path-generation algorithm receives information about the vehicle and the environment to plan a desired path 
$$\bm{p_d}(\gamma) = \sum_{j=0}^N \bm{\bar{p}_{j,N}} b_{j,N}(\gamma) \, , \qquad \gamma \in [0, T],$$
where $\gamma$ is an independent variable referred to as \textit{virtual time}, $\bm{\bar{p}_{j,N}}$, $j = 0, \ldots , N$ are polynomial coefficients, and $b_{j,N}(\cdot)$, $j = 0, \ldots , N$ are polynomial basis functions of order $N$;
$(ii)$ the computed path is passed to the path-following algorithm. The algorithm generates yaw commands, $\psi_c$, to track the \textit{virtual target} $\bm{p}_d(\gamma)$. The virtual time $\gamma$ is also regulated by the path-following algorithm; $(iii)$ the inner-loop controller determines the control surface inputs necessary to execute the yaw commands from the path-following controller. 

On the other hand, the trajectory-tracking architecture depicted in Figure~\ref{fig:tt_block_diagram} operates as follows: $(i)$ the trajectory-generation algorithm receives information about the vehicle and the environment to plan a desired trajectory 
$$\bm{p_d}(t) = \sum_{j=0}^N \bm{\bar{p}_{j,N}} b_{j,N}(t) \, , \qquad t \in [0, T];$$
$(ii)$ the computed trajectory is passed to the trajectory-tracking algorithm, which generates velocity and yaw commands, i.e., $v_c$ and $\psi_c$, respectively, to track the trajectory $\bm{p}_d(t)$; $(iii)$ the inner-loop controller determines the control surface inputs necessary to execute the velocity and yaw commands from the trajectory-tracking controller. 


These control architectures have been developed considering the hazardous operating conditions \textcolor{red}{underwater vehicles} are often subject to in their missions, so the performance bounds have been rigorously considered to avoid potential obstacles or unwanted interaction with the surface.

The motion planning strategy employed in this architecture uses Bernstein polynomial approximation as a means of generating a continuous path, $\bm{p_d}(\cdot)$, or trajectory, $\bm{p_d}(t)$, from a set of defined control points \cite{cichella2020optimal,kielas2019bebot,kielas2022bernstein}. Optimization techniques demonstrated in \cite{cichella2019optimal,cichella2022consistency} can be used to generate approximately optimal paths/trajectories for the vehicle, while constraining limits to parameters like vehicle speed and angular rate.

In this section, we delve into the specifics of the path-following and trajectory-tracking control architectures. Additionally, we will provide further insight into the inner-loop controller design.

\subsection{Path Following}
The path-following algorithm used to follow the generated trajectory draws from the high-level motion-control algorithm outlined in \cite{cichella2011geometric}, \cite{kaminer2017time}, \cite{cichella20113d}, \cite{cichella20133dmultirotor}. 
Consider a geometric path $\bm{p_d}:[0,T_f] \to \mathbb{R}^2$ defined in an inertial frame, $\mathcal{I}$, and parameterized by virtual time, $\gamma:\mathds{R}^+\rightarrow[0,T_f]$. The dynamics of the virtual time can be used as an extra degree of freedom to move the virtual target along the path as a function of the vehicles state. Let the parallel transport frame \citep{kaminer2017time}, denoted as $\mathcal{T}$, define the orientation of the virtual target, $\bm{p_d}(\gamma)$. The $\mathcal{T}$ frame’s orientation with respect to $\mathcal{I}$ is given by the rotation matrix $\bm{R}_T^I(\gamma) \triangleq [\bm{\hat t}_1(\gamma),\bm{\hat t}_2(\gamma)]$, with $\bm{\hat t}_1(\cdot)$ being a unit vector tangent to the velocity of the path, thus $\bm{\hat{t}}_1(\gamma)=\bm{p_d}'(\gamma)/||\bm{p_d}'(\gamma)||$, where $\bm{p_d}'(\gamma) = d\bm{p_d}(\gamma)/d\gamma$. The vector $\bm{\hat t}_2(\gamma)$ is orthonormal to $\bm{\hat t}_1(\gamma)$ and found considering curvature and torsion of the path at $\gamma$. The angular velocity of this frame with respect to $\mathcal{I}$ is denoted by $\bm{\omega}_T$.

\begin{figure}
    \centering
    \includegraphics[width=0.49\textwidth]{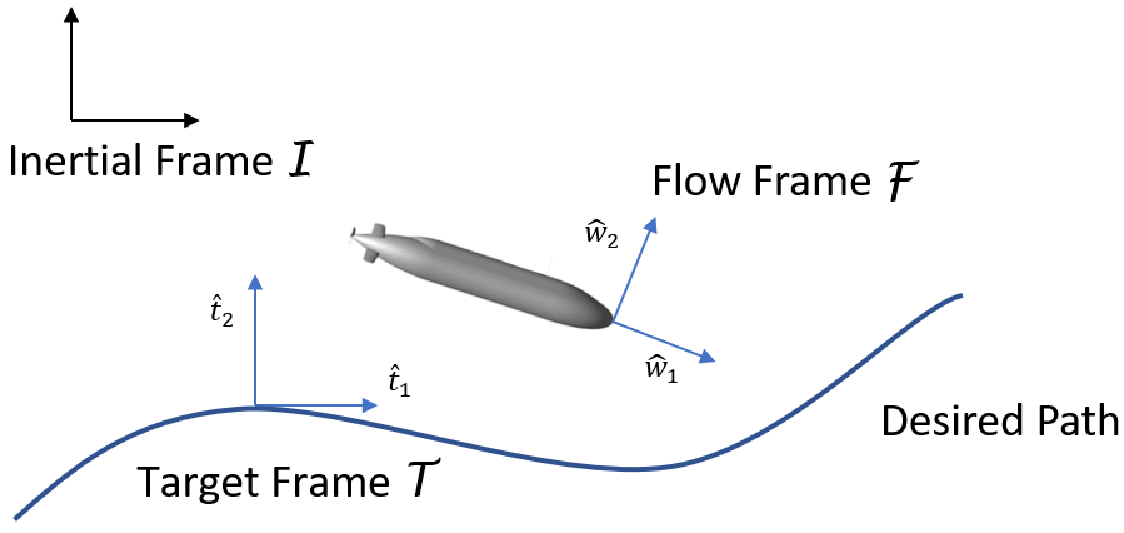}
    \caption{Geometry associated with the path-following and trajectory tracking problems}
    \label{fig:vector_summary}
\end{figure}

Consider the flow frame, denoted as $\mathcal{W}$, which has its origin, $\bm{p}$, at the vehicle center of mass. Rotation matrix $\bm{R}_W^I \triangleq [\bm{\hat w}_1,\bm{\hat w}_2] $
gives the frame's orientation with the $x$-axis being aligned to the vehicles velocity. In other words, given the flow velocity vector $\bm{v}_W=[v,0]^\top$, the vehicle's kinematics are governed by 
\begin{equation} \label{EquationPFvehicledyn}
\dot{\bm{p}} = \bm{R}_W^I\bm{v}_W
\end{equation}
As will become clear later, the vehicle's orientation, i.e., matrix $\bm{R}_W^I$, is adjusted to guarantee convergence of the vehicle to the desired path. Fig.~\ref{fig:vector_summary} provides a visual aid for the discussed frames and vectors. To this end, let us define the path-following error as follows
\begin{equation} \label{eq:poserror}
    \bm{p_T}=\bm{R_I^T}(\bm{p}-\bm{p_d}(\gamma)) \triangleq [x_T,y_T]^T
\end{equation}
with dynamics 
\begin{equation} \label{eq:positionerrordynamics}
    \begin{aligned}
    \dot{\bm{p_T}}   &=
    \dot{\bm{R}}_{I}^{T}(\bm{p}-\bm{p_d}(\gamma)) + {\bm{R}}_{I}^{T} \dot{\bm{p}} - {\bm{R}}_{I}^{T} \dot{\bm{p}}_{\bm{d}} \\ 
    &= -\bm{\omega_T}\times \bm{p_T}+\bm{R_W^T}
    \begin{bmatrix}
    v \\
    0 
    \end{bmatrix}
    -
    \begin{bmatrix}
    ||\bm{p_d}'(\gamma)||\dot{\gamma} \\
    0
    \end{bmatrix}
    \end{aligned}
\end{equation}

\hspace{1mm}

With the above setup, the path-following problem can be defined as follows.
\begin{prob}
{Consider a vehicle with dynamics governed by Equation \eqref{EquationPFvehicledyn}. Let the vehicle be equipped with an autopilot that provides tracking capabilities of orientation (yaw) commands, i.e., $\bm{R}_c$. In other words, 
\begin{equation} \label{asm:AP}
\bm{R}_c \equiv \bm{R}_W^I 
\end{equation}
Derive control laws for $R_c$ and for the rate of progression of the virtual time, $\dot{\gamma}$, such that the path-following position error defined by Equation \eqref{eq:poserror}, with dynamics governed by Equation \eqref{eq:positionerrordynamics}, converges to a neighborhood of the origin.}
\end{prob}
To solve the path-following problem, we let the dynamics of the virtual time be governed by
\begin{equation} \label{eq:gammad}
	\dot{\gamma} = \frac{\left[{v}\bm{\hat{w}_1}+k_\gamma (\bm{p}-\bm{p_d}(\gamma))\right]^\top		\bm{\hat{t}}_{1}(\gamma)}{\left\Vert\bm{p_d'}(\gamma)\right\Vert} \, 
\end{equation} 
and the orientation command be given by
\begin{equation} \label{eq:pfrot}
    R_c =  R_T^I
    \begin{bmatrix}
    \frac{d}{(d^2+y_T^2)^{1/2}} & \frac{y_T}{(d^2+y_T^2)^{1/2}} \\
    \frac{-y_T}{(d^2+y_T^2)^{1/2}} & \frac{d}{(d^2+y_T^2)^{1/2}} 
    \end{bmatrix}
\end{equation}
{In the equations above, $d,k_\gamma>0$ are control design parameters.}

\begin{thm} \label{thm:PF}
Consider a vehicle equipped with an inner-loop autopilot that satisfies Equation \eqref{asm:AP}.
There exist control parameters $d$ and $k_\gamma$ such that, for any initial state $p_T(0)$ the rate of progression of the virtual time \eqref{eq:gammad} and the orientation command  \eqref{eq:pfrot} ensure that the path-following position error 
$p_T(t)$ is asymptotically stable.
\end{thm}
\begin{prf}
{The proof of Theorem \ref{thm:PF} is provided in the Appendix, Section \ref{sec:appendix_pf}.}
\end{prf}

\subsection{Trajectory Tracking}

Here the trajectory tracking algorithm is presented as an alternative to the path-following algorithm.
The primary difference between the trajectory tracking and path-following algorithms is that trajectory tracking is constrained by time, while path following is not. Due to this constraint, the vehicle will arrive at the final destination at the exact final time prescribed by the user.

Now we define $\bm{R_T^I}(t)$ as the matrix representation of the orientation of frame $\mathcal{T}$ with respect to frame $\mathcal{I}$. This orientation is characterized by $\bm{R_T^I}(t)=[\bm{\hat{t}_1}(t),\bm{\hat{t}_2}(t)]$ such that the velocity of the target $\bm{v_d}= [v_d,0]^T$, where $v_d$ is the target's speed.  

Similarly to the path-following approach, the position $\bm{p}$ of the vehicle (the origin of the flow frame $\mathcal{W}$) is governed by \eqref{EquationPFvehicledyn}, The position error is now defined as
\begin{equation} \label{eq:ttpositionerror}
    \bm{e}_p=\bm{p}_d(t)-\bm{p},
\end{equation}
where $\bm{p}_d(t)$ is the position of the target with respect to $\mathcal{I}$ resolved in $\mathcal{I}$. Notice that the equation above emphasizes the dependency of the desired trajectory on $t$, which is in contrast to the path-following approach, where the desired position is expressed as a function of virtual time.
The dynamics of the position error is defined as
\begin{equation} \label{eq:ttpositionerrordyn}
    \bm{\dot{e}}_p= \bm{R}_T^I
    \begin{bmatrix}
    v_d \\
    0 
    \end{bmatrix} -
    \bm{R}_W^I
    \begin{bmatrix}
    v \\
    0 
    \end{bmatrix}.
\end{equation}

With this setup, the trajectory tracking problem is defined as follows.
\begin{prob}
Consider a vehicle with dynamics governed by Equation \eqref{EquationPFvehicledyn}. Let the vehicle be equipped with an autopilot that provides tracking capabilities of feasible speed and orientation (yaw) commands, i.e., $v_c$ and $\bm{R}_c$, respectively. In other words, 
\begin{equation} \label{asm:APTT}
v_c \equiv v \, \quad \bm{R}_c \equiv \bm{R}_W^I 
\end{equation}
Derive control laws for $v_c$ and $R_c$ such that the path-following position error defined by Equation \eqref{eq:ttpositionerror}, with dynamics governed by Equation \eqref{eq:ttpositionerrordyn}, converges to a neighborhood of the origin. 
\end{prob}
To solve the problem above the following control laws are proposed for the velocity and orientation of the vehicle
\begin{equation} \label{eq:ttvelocity}
v_c = (k_p\bm{e}_p+v_d\textcolor{red}{\bm{\hat{t}_1}})^T\textcolor{red}{\bm{\hat{w}_1}}.
\end{equation}
\begin{equation} \label{eq:ttrot}
    R_c = [\bar{\bm{b}}_{1D}, \bar{\bm{b}}_{2D}] , \qquad \bar{\bm{b}}_{1D} = \frac{k_p \bm{e}_p + v_d \bm{\hat{t}_1}}{||k_p \bm{e}_p + v_d \bm{\hat{t}_1}||} 
\end{equation}
{where $k_p$ is a control gain and the vector $\bar{\bm{b}}_{2D}$ is chosen to be orthonormal to $\bar{\bm{b}}_{1D}$.

\begin{thm}\label{thm:TT}
{
Consider a vehicle equipped with an inner-loop autopilot that satisfies Equation \eqref{asm:APTT}. There exists control parameters $d$ and $k_p$ such that, for any initial state $e_p(0)$ the velocity and orientation commands, $v_c$ and $R_c$, respectively, ensure that position error 
$e_p(t)$ is asymptotically stable.}
\end{thm}
\begin{prf} 
{The proof of Theorem \ref{thm:TT} is provided in the Appendix, Section \ref{sec:appendix_tt}.}
\end{prf}

\subsection{Adaptive Inner-Loop Control}
With the rotation command determined by either the path-following controller or the trajectory tracking controller, the inner loop controller ensures that this command is properly carried out. Assuming that the submarine is already equipped with an autopilot which can follow input depth and yaw reference signals by manipulating $\delta_V$ and $\delta_H$ as in Equation \eqref{CommandsVH}, an $\mathcal{L}_1$ augmentation is formulated for it in order to improve performance. \textcolor{red}{Similar $\mathcal{L}_1$ augmentation can be seen in \cite{kaminer2010path} and \cite{cao2007stabilization}.}

In this paper we assume that the closed-loop system 
 consisting of the vehicle and its autopilot is given by
\begin{equation} \label{eq:APsys}
    \bm{\mathcal{G}_p}(s)
    \begin{cases}
        \dot{\bm{x}}(t) \!=\! \bm{A_{p}} \bm{x}(t) \!+\! \bm{B_{p}} (\bm{u_{ad}}(t)\!+\!\bm{f}(t,\bm{x}(t)))\\
        \bm{y}(t) \!=\! \bm{C_{p} x}(t),  \\
    \end{cases}
\end{equation}
where $\bm{u_{ad}}(t){= [\psi_{ad}(t),z_{ad}(t)]^T}$ is the input yaw and depth reference signals, the output $\bm{y}(t)   = [\psi(t),z(t)]^T$ is the actual vehicle yaw angle and depth, and $\bm{f}(t,\bm{x}(t))$ is a time-varying function capturing system uncertainties and external disturbances. $\left\{\bm{A_{p}},\;\bm{B_{p}},\;\bm{C_{p}}\right\}$ is a controllable-observable triple describing the system. The proposed formulation and design of the adaptive control system are explicitly tailored to handle the dynamics associated with yaw and depth control. In this context, speed control is intentionally omitted. The dynamics pertaining to speed control tend to exhibit greater simplicity when compared to those involved in yaw and depth control. Speed control for the vehicle under consideration primarily revolves around the regulation of the propulsion system or the adjustment of thrust generated by the propellers. These dynamics lend themselves to the implementation of relatively simpler control strategies.

We now introduce desired system
\begin{equation} \label{eq:desiredM}
 \bm{M}(s) \triangleq \bm{C_{m}}\left(s\mathds{I}-\bm{A_{m}}\right)^{-1}\bm{B_{m}}
\end{equation}
as a design parameter of the $\mathcal{L}_1$ controller, specifying the desired depth and yaw behavior. The system must be selected such that $\bm{A_m}$ is Hurwitz, $\bm{C_m}\bm{B_m}$ is nonsingular, and $\bm{M}(s)$ does not have a non-minimum-phase transmission zero, see \cite{jafarnejadsani2019l1}. The dynamics of this desired system can be written as
\begin{equation}
\label{eq:desired}
    \bm{y_m}(s) = \bm{M}(s) \bm{K_{g}} \bm{u_{ref}}(s), \quad \bm{K_{g}} = -(\bm{C_{m}} \bm{A_{m}}^{-1} \bm{B_{m}})^{-1} ,
\end{equation}
{where $u_{ref}(s)=[\psi_{c}(s),z_{c}(s)]^\top$ are Laplace transforms of the reference commands to be tracked (given by the motion planner, path-following or trajectory-tracking controller), and $\bm{y_m}$ is the desired output.} In the case of path following or trajectory tracking, the yaw command for the vehicle can be obtained from the 2D rotation matrix, Equations \eqref{eq:pfrot} or \eqref{eq:ttrot}. The objective of the adaptive controller is to design $\bm{u}_{ad}$ so that output $\bm{y}$ of \eqref{eq:APsys} tracks output $\bm{y}_m$ of \eqref{eq:desired}.

The adaptation law of this controller provides a discrete-time estimate $\bm{\hat{\sigma}_d}(t)$ for the unknown function $\bm{f}(t,\bm{x}(t))$ in Equation \eqref{eq:APsys}, and is given by
\begin{equation}
\label{eq:adaptation_law_}
    \bm{\hat{\sigma}_d}[i] = - \bm{\Phi}^{-1}(T_s)e^{\bm{\Lambda} \bm{A_{m}} \bm{\Lambda}^{-1}T_s}\bm{1}_{n_m2}(\bm{\hat{y}_d}[i]-\bm{y_d}[i])
\end{equation}
where
$
\bm{\hat{\sigma}_d}(t) = \bm{\hat{\sigma}_d}[i],\quad t \in [iT_s,(i+1)T_s),\quad i \in \mathds{Z}_{\geq 0}
$,
$y_d[i]=y(iT_s)$ is the sampled output, 
\[
\Phi (T_s) = \int_0^{T_s} e^{\Lambda A_m \Lambda^{-1} (T_s-\tau)} \Lambda d\tau ,
\]
and 
\[
\Lambda = 
\begin{bmatrix} 
C_m \\ D \sqrt{P} 
\end{bmatrix} ,
\]
where $P$ is matrix solution to $A_{m}^\top P+P A_{m}=-Q$ for a given positive definite matrix $Q$, $P = \sqrt{P}^\top \sqrt{P}$, and $D$ is a matrix that satisfies
$
D\left( C_m \left(\sqrt{P}\right)^{-1} \right)^{\top} = 0 .
$
With this estimation in place, a discrete-time output predictor can be formulated as
\begin{equation}
\label{eq:output_predictor}
\begin{split} 
    \bm{\hat{x}_d}[i+1] & = e^{\bm{A_{m}}T_s}\bm{\hat{x}_d}[i] \\ &  +\bm{A_{m}}^{-1}\left(e^{\bm{A_{m}} T_s}-\mathds{I}_{n_m}\right)\left(\bm{B_{m}} \bm{u_d}[i]+\bm{\hat{\sigma}_d}[i]\right), \\   \bm{\hat{y}_d}[i] & = \bm{C_{m}}\bm{\hat{x}_d}[i],\quad \bm{x_d}[0] = \bm{C_{m}}^{\dagger}\bm{y}_0
\end{split}
\end{equation}
which replicates the desired closed-loop dynamics of the system given by Equation \eqref{eq:desired}.

Finally, the control law giving the input $\bm{u_{ad}}$ is given as
\begin{equation*}
    \bm{u_{ad}}(t) = \bm{u_d}[i],\quad t \in [iT_s,(i+1)T_s),\quad i \in \mathds{Z}_{\geq 0}
\end{equation*}
where
\begin{equation}
\label{eq:control_law}
\begin{split}
    \bm{x_u}[i+1] & = e^{\bm{A_o}T_s}\bm{x_u}[i] \\ & +\bm{A_o}^{-1}\left(e^{\bm{A_o} T_s}-\mathds{I}_{n_o}\right)\left(\bm{B_o}e^{-\bm{A_{m}}T_s}\bm{\hat{\sigma}_d}[i]\right), \\
    \bm{u_d}[i] & = \bm{K_{g}}\bm{u_{ref}}[i]-\bm{C_o}\bm{x_u}[i],\quad \bm{x_u}[0] = 0
\end{split}
\end{equation}
Here, the triple $\left\{\bm{A_{o}},\;\bm{B_{o}},\;\bm{C_{o}}\right\}$ is the minimal state-space realization of the transfer function
\begin{equation}
\label{eq:filter}
    \bm{O}(s) = \bm{C}(s)\bm{M}^{-1}(s)\bm{C_{m}}\left(s\mathds{I}_{n_o}-\bm{A_{m}}\right)^{-1}
\end{equation}
and $\bm{C}(s)$ is a strictly proper stable transfer function such that $\bm{C}(0) = \mathds{I}_2$. \cite{hovakimyan2010} and \cite{jafarnejadsani2018robust} should be consulted for discussion of the bounds \textcolor{red}{that define the robustness} of this controller and for \textcolor{red}{additional} information on parameter tuning.

\section{Results} \label{sec:results}

\subsection{Adaptation}
\textcolor{red}{A direct comparison is made between the adaptive inner loop controller presented in this work and the inner loop controller designed in \cite{overpelt2015free} for the vehicle in question.} To compare the performance of the adaptive and non-adaptive inner-loop controllers, a simple near surface depth-keeping maneuver is analyzed. This maneuver is chosen because it highlights the surface suction effect, which introduces a large disturbance in the depth dynamics. We assume that the vehicle is equipped with an autopilot consisting of two proportional-derivative (PD) controllers with gains of $k_{p_z}=3$ and $k_{d_z}=3$ for the depth controller, and gains of $k_{p_\psi}=3$ and $k_{d_\psi}=12.2$ for the heading controller. \textcolor{red}{This autopilot was developed and tuned by the authors in \cite{overpelt2015free}, and is also presented in \cite{carrica2016cfd} and \cite{carrica2019near}}. I.e., the PD controllers are formulated below
\[\delta_V=k_{p_z}(z_{ad}-z)+k_{d_z}(\dot{z}_{ad}-\dot{z})\]
\[\delta_H=k_{p_\psi}(\psi_{ad}-\psi)+k_{d_\psi}(\dot{\psi}_{ad}-\dot{\psi})\]
where $z_{ad}$ and $\psi_{ad}$ are the augmented reference signals from the adaptive controllers and with each stern plane angle $\delta_1....\delta_5$ being derived from Equation~\ref{CommandsVH}.

Next, we design the adaptive controller which regulates the reference $z_{ad}$ and $\psi_{ad}$ to be tracked by the autopilot. To this end, the desired system introduced in Equation \eqref{eq:desiredM} is designed as follows:
\begin{equation}\label{eq:secondordertf}
    \bm{M}(s) = \begin{bmatrix} \frac{\omega_n^2}{s^2+2\zeta\omega_n s+\omega_n^2} & 0 \\ 0 & \frac{\omega_n^2}{s^2+2\zeta\omega_n s+\omega_n^2} 
\end{bmatrix} \end{equation}
with $\omega_n = 0.08$ and $\zeta = \textcolor{red}{1}$. \textcolor{red}{The filter $\bm{C}(s)$ in Equation \eqref{eq:filter} must be strictly proper stable and with static gain \(C(0)=1\) \cite{hovakimyan2010}. Motivated by \cite{pettersson2012analysis,li2007optimization}, the filter is designed to allow frequencies above the bandwidth of the closed-loop reference system response while simultaneously blocking frequencies exceeding the effective control bandwidth. After extensive tuning through simulation experiments, we have formulated a second-order low-pass filter with a cutoff frequency of \(\omega_c = 1.5 \omega_n\) and a damping coefficient \(\zeta = 1\), i.e., 
\begin{equation}\label{eq:lowpassfilter}
    \bm{C}(s) = \begin{bmatrix} \frac{\omega_c^2}{(s+\omega_c)^2} & 0 \\ 0 & \frac{\omega_c^2}{(s+\omega_c)^2}  \end{bmatrix}
\end{equation}}

The maneuver shown in Fig.~\ref{fig:near_surface} has an initial and desired depth of -15 m. It can be seen that the adaptive controller is capable of estimating the unknown disturbance, and augmenting the reference signal sent to the autopilot to counteract the disturbance, as with the augmented reference, the vehicle is capable of converging to the desired depth.

\begin{figure}
    \centering
    \includegraphics[width=.52\textwidth]{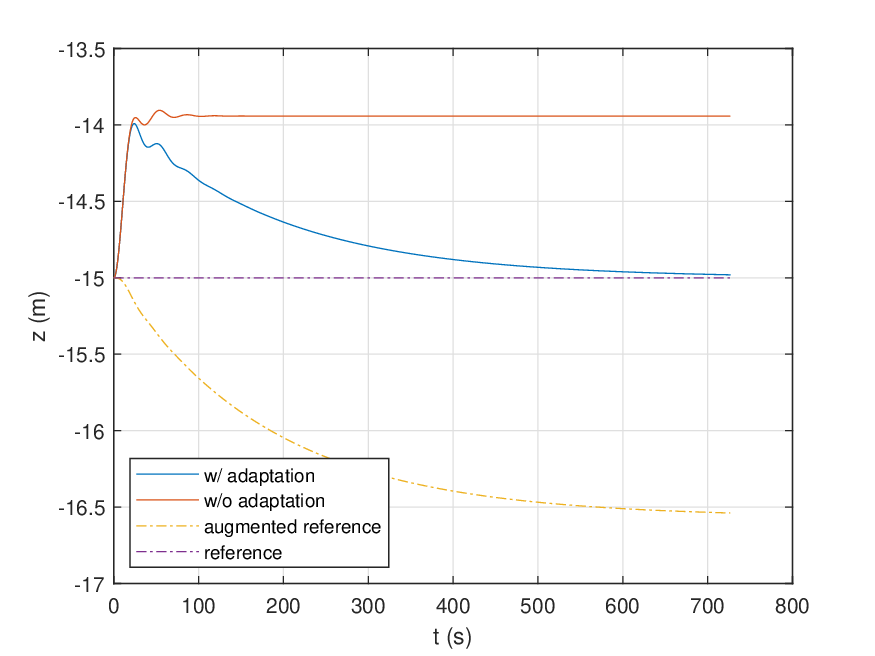}
    \caption{Near surface depth-keeping maneuver. The initial and desired position in this case is -15 m}
    \label{fig:near_surface}
\end{figure}

\textcolor{red}{To further compare the performance when using the adaptive controller to augment the commercial autopilot and the unaugmented autopilot, a turning circle maneuver is presented at two depths. These maneuvers show that the adaptive controller is able to estimate unknown disturbances for a more complex maneuver. A turning circle maneuver at depth and near the surface is shown in Fig. \ref{fig:tc_100m} and Fig. \ref{fig:tc_15m} respectively. While the commercial autopilot may be satisfactory at depth, it is evident that for near surface maneuvers where suction effects are more prominent, adaptive augmentation can accurately estimate the disturbance and correct course.}

\begin{figure}
    \centering
    \includegraphics[width=.52\textwidth]{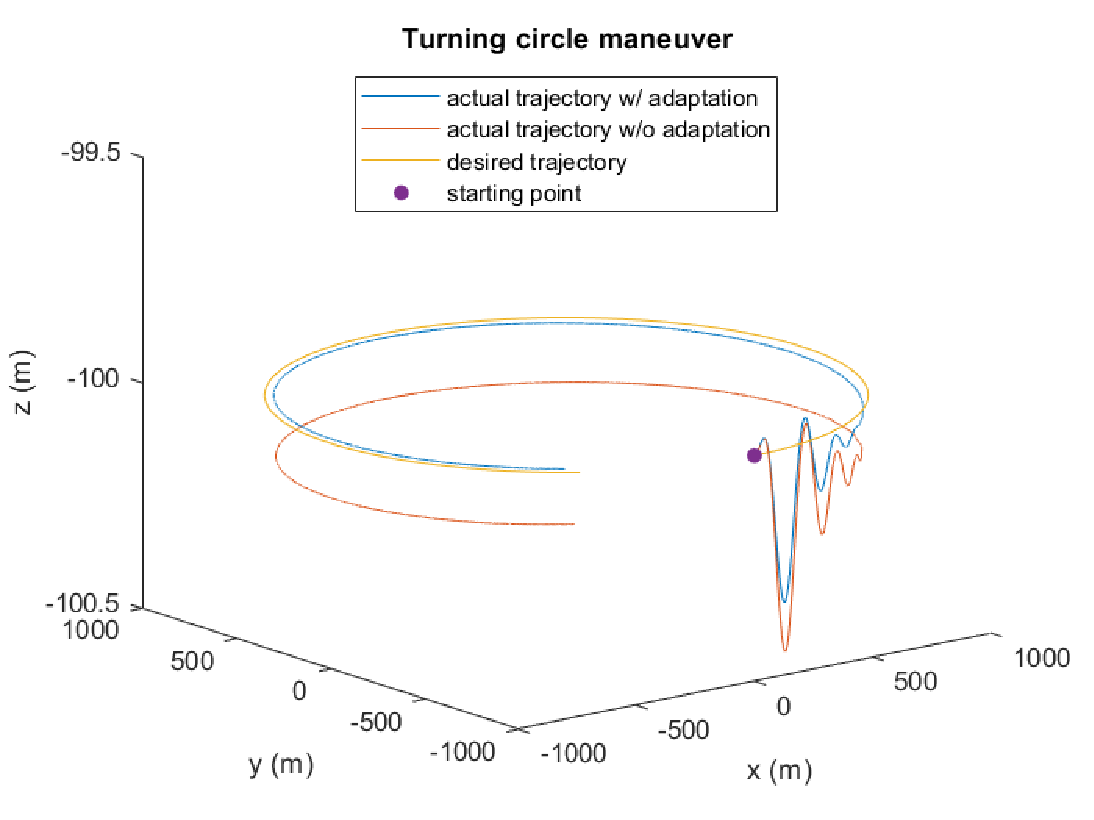}
    \caption{Turning circle maneuver at a depth of 100m}
    \label{fig:tc_100m}
\end{figure}

\begin{figure}
    \centering
    \includegraphics[width=.52\textwidth]{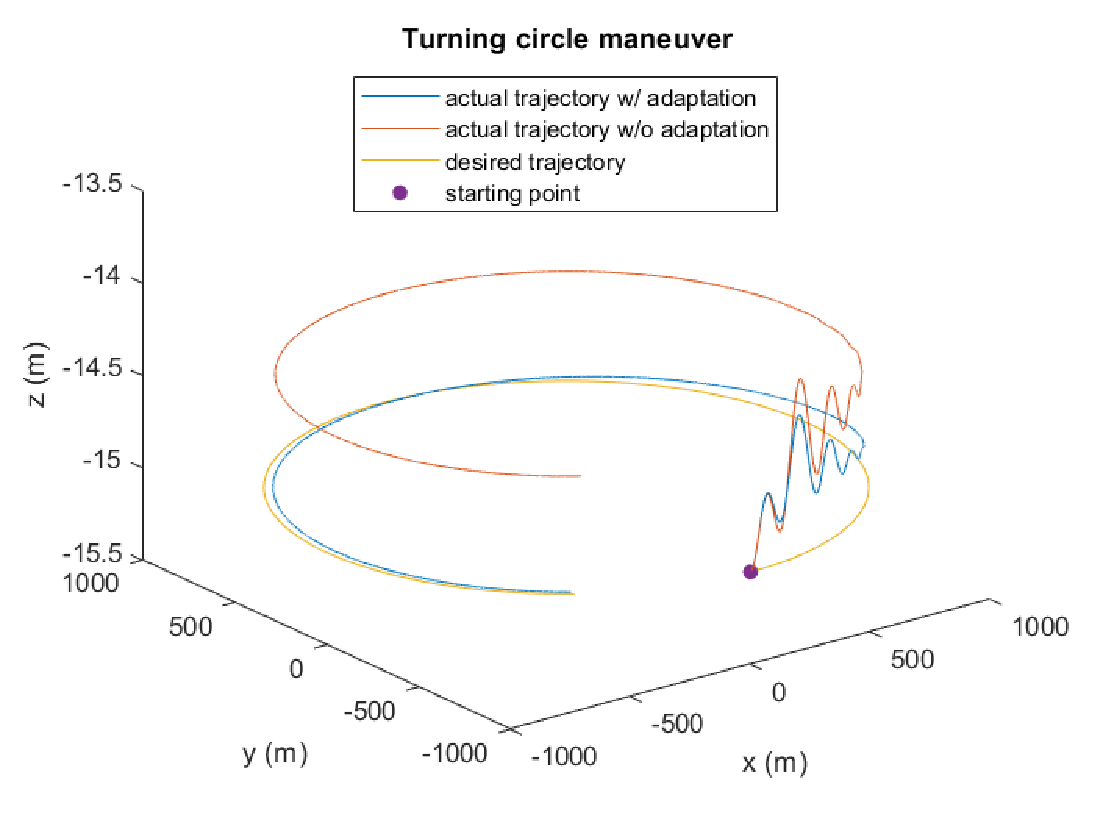}
    \caption{Turning circle maneuver at a depth of 15m}
    \label{fig:tc_15m}
\end{figure}

\subsection{Path-Following and Trajectory Tracking Bathymetry Example}
An example of maneuvering following a pre-established trajectory near complex bathymetry is presented in this section. 
The objective of this calculation is to demonstrate the ability of the path-following and trajectory-tracking controllers to track a complex path/trajectory using the ROM, as well as to provide an opportunity to compare both control methods. Fig.~\ref{fig:canyon_geo} illustrates the example at hand. The path is set to approximately follow the thalweg (the line of lowest elevation at each cross-section within a valley) of the canyons maintaining a distance of roughly 50 m between the CG and the bottom. The desired path to be tracked by the path-following controller is defined as
$$\bm{p_d}(\gamma) = \sum_{j=0}^N \bm{\bar{p}_{j,N}} b_{j,N}(\gamma) \, , \qquad \gamma \in [0, T]$$
where the coefficients ${\bm{\bar{p}_{j,N}}}$, $j=0,\ldots,N$, are pre-computed off-line, $b_{j,N}(\gamma)$ is the Bernstein polynomials basis, and $\gamma$ is governed by Equation \eqref{eq:gammad}. Similarly, the trajectory-tracking controller is tasked to track trajectory
$$\bm{p_d}(t) = \sum_{j=0}^N \bm{\bar{p}_{j,N}} b_{j,N}(t) \, , \qquad t \in [0, T]$$
In both cases, the final time is set to $T=500$s, which corresponds to a travelled distance of roughly 1.5 miles from the beginning to end location at approximately $8$kts. The path-following controller, being unconstrained by time, achieves the maneuver in approximately $400$s at a desired speed of approximately $10$kts. The control gains are defined as $d=\textcolor{red}{50}$, $k_\gamma=1$, $k_p=0.1$ for the path-following and trajectory-tracking examples. As the vehicle reaches the transition from the La Jolla canyon to the Scripps canyon, shown in Fig.~\ref{fig:canyon_geo}, a significant horizontal command is required, as shown in Fig.~\ref{fig:scripps_cmd}(a) and Fig.~\ref{fig:scripps_cmd}(b) at roughly $125$s for the path-following example, and $165$s for the trajectory tracking example. \textcolor{red}{This sudden, significant horizontal command is beyond the dynamics of the vehicle, introducing horizontal position error, as seen in} Figure \ref{fig:scripps_pos_error}. Even with this limitation, both the path-following and trajectory-tracking controllers allow the vehicle to accomplish the maneuver.


The trajectory of the vehicle is of course determined by the actuation of the control surfaces, and it is not surprising that the controllers impose similar command histories for each method, as shown in Fig.~\ref{fig:scripps_cmd}. Note that due to the combined effect of vertical and horizontal commands on the stern planes, the errors in tracking can be very different even though the commands are similar. {As the vehicle transitions between the two canyons, the virtual time slows significantly, as shown in Fig.~\ref{fig:scripps_gammadot}. The virtual time slows down in this case because this section of the maneuver {requires more control authority than present, introducing some path error which is then procedurally reduced by adjusting the rate of the virtual time.} Additionally, the \textcolor{red}{actual and desired headings} for both controllers are shown in Fig.~\ref{fig:scripps_psiref}, and the commanded velocity for the trajectory tracking controller is shown in Fig.~\ref{fig:scripps_vel_cmd}.}

Videos showing a simulation of this bathymetry maneuver for each controller, created using MATLAB Simscape\texttrademark, are also shown on our github web page. \footnote{\url{https://github.com/caslabuiowa/IowaBB2model}}

\begin{figure}
  \centering
    \includegraphics[width=.5\textwidth]{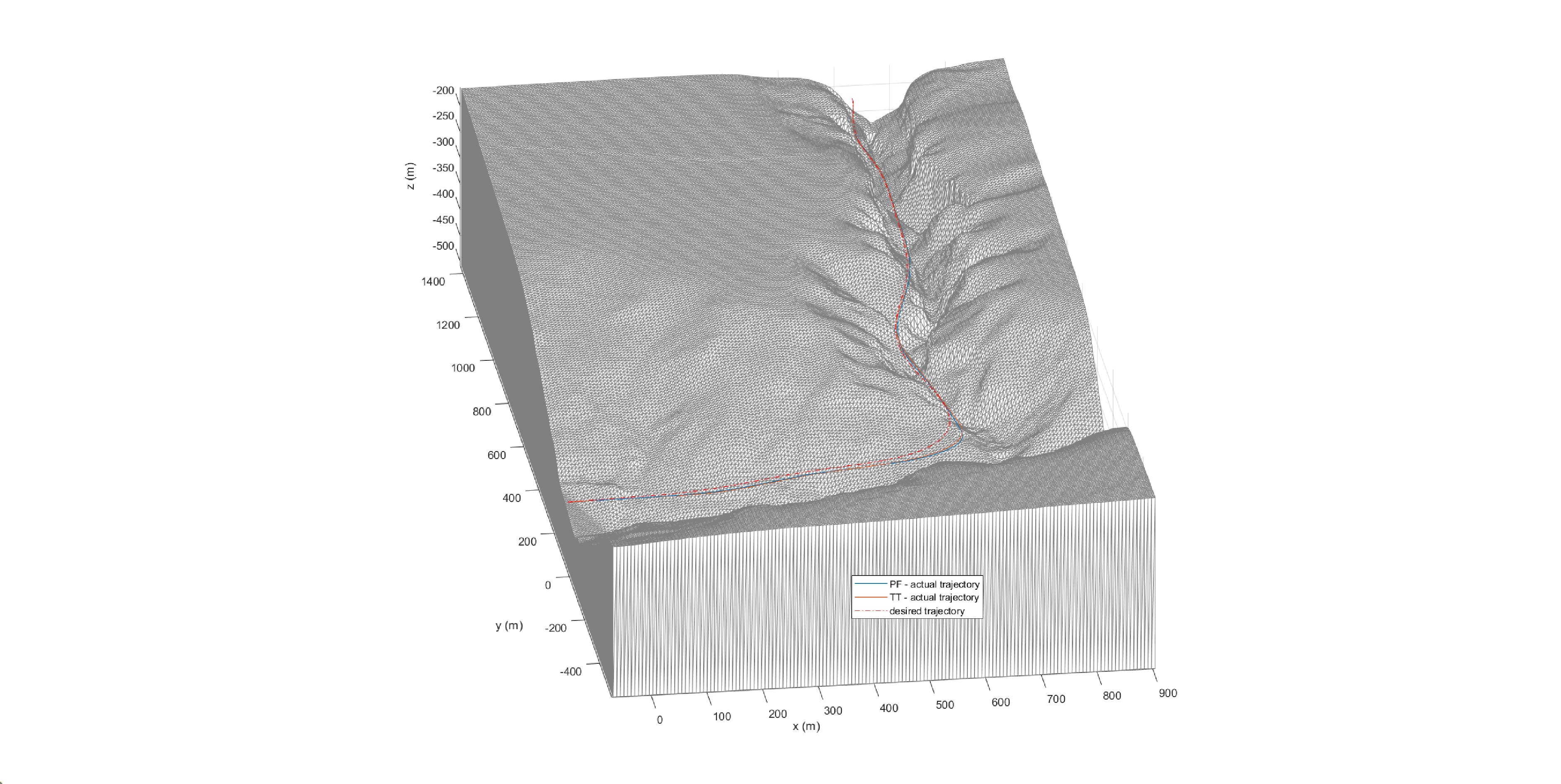}
    \caption{Scripps and La Jolla canyons with prescribed path $\bm{p_d}$, actual trajectory, and actual path.}
    \label{fig:canyon_geo}
\end{figure}

\begin{figure}
  \centering
    \includegraphics[width=.52\textwidth]{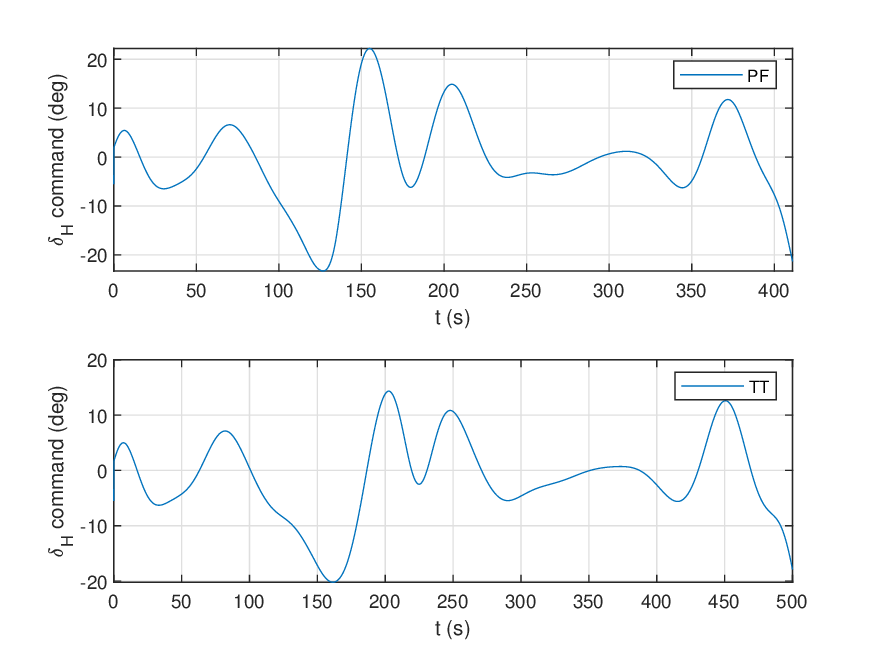}
    \includegraphics[width=.52\textwidth]{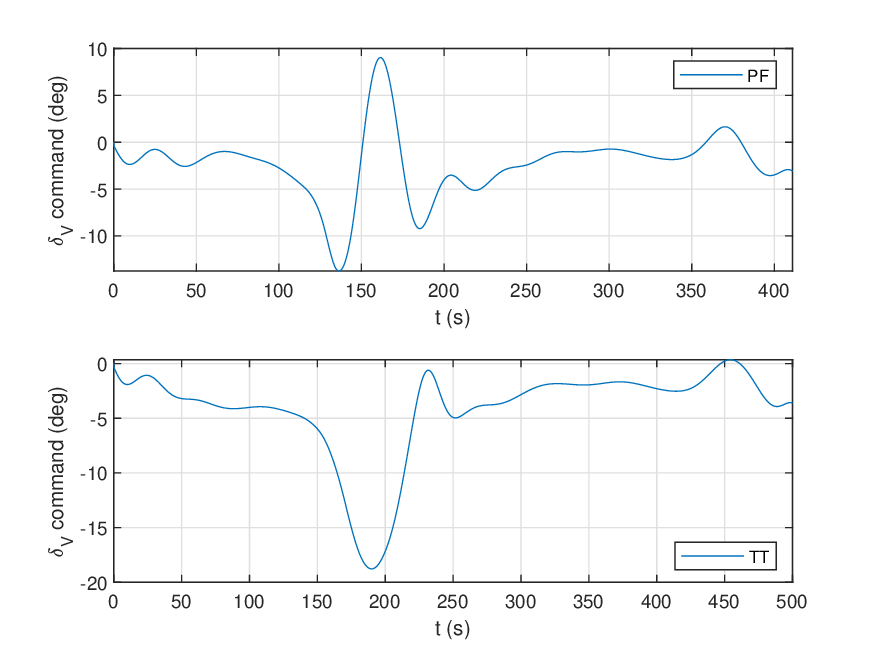}
    \caption{Control plane commands for the Scripps and La Jolla maneuver. Top: horizontal command; bottom: vertical command}
    \label{fig:scripps_cmd}
\end{figure}

\begin{figure}
  \centering
    \includegraphics[width=.52\textwidth]{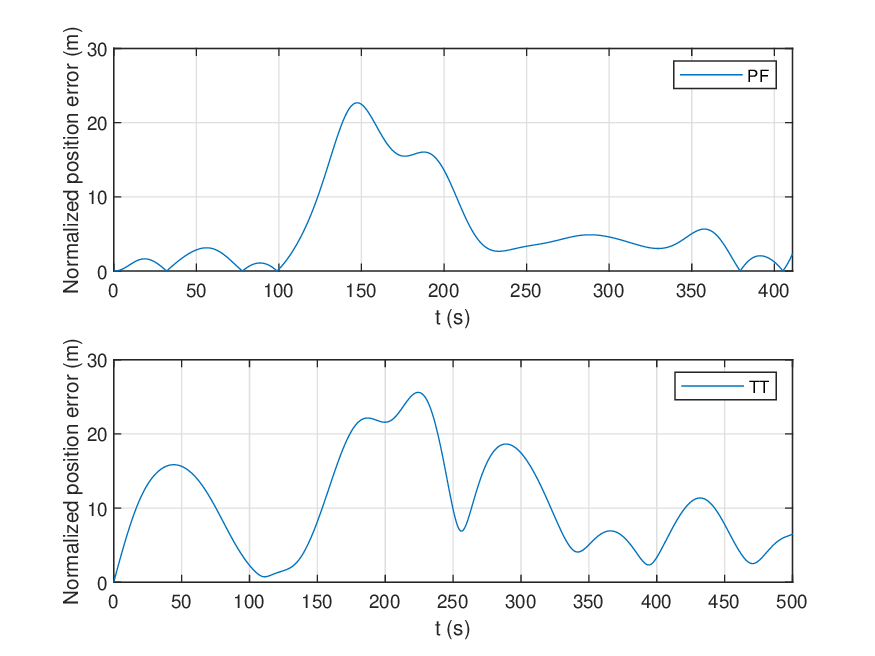}
    \caption{Normalized horizontal position error (x and y) between target and path/trajectory for the Scripps and La Jolla maneuver}
    \label{fig:scripps_pos_error}
\end{figure}

\begin{figure}
  \centering
    \includegraphics[width=.52\textwidth]{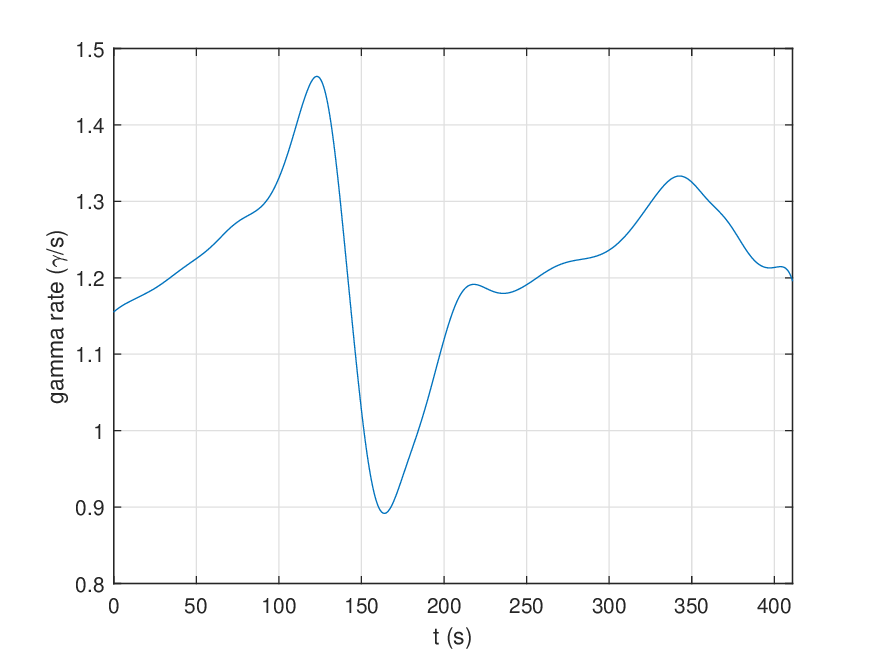}
    \caption{Rate of the virtual time $\gamma$ for the Scripps and La Jolla maneuver}
    \label{fig:scripps_gammadot}
\end{figure}

\begin{figure}
  \centering
    \includegraphics[width=.52\textwidth]{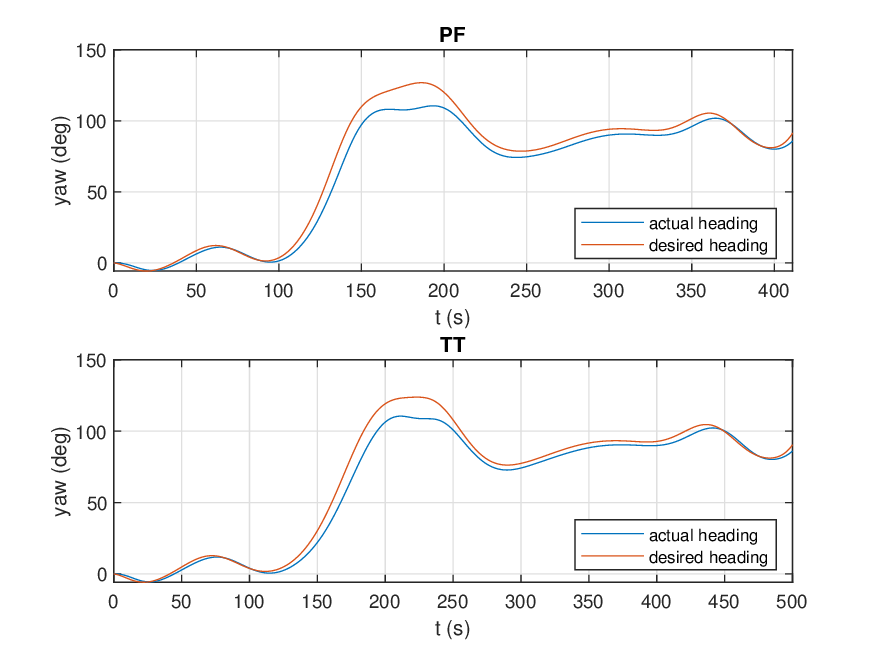}
    \caption{Actual and desired heading for the Scripps and La Jolla maneuver}
    \label{fig:scripps_psiref}
\end{figure}

\begin{figure}
  \centering
    \includegraphics[width=.52\textwidth]{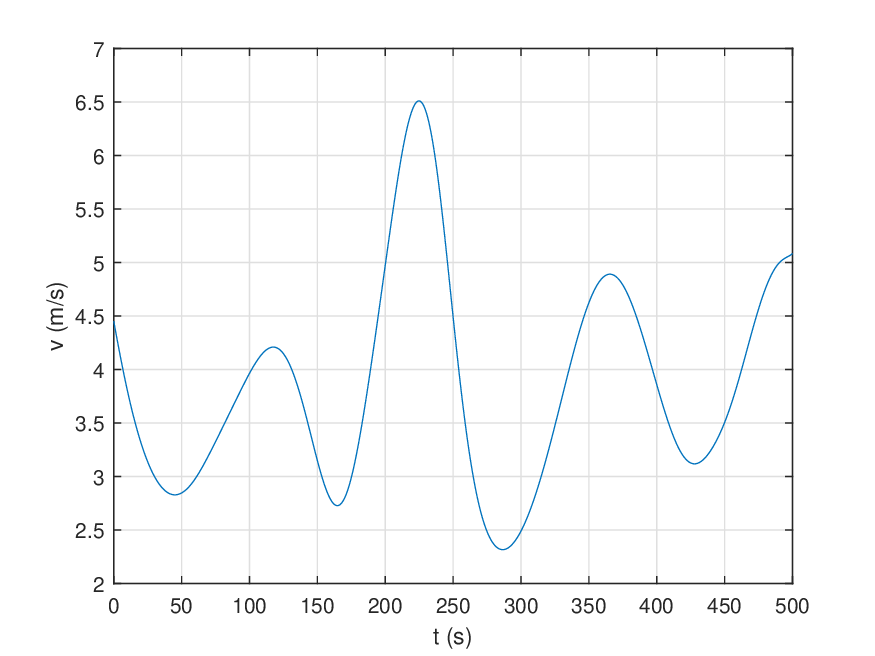}
    \caption{Velocity command history of the trajectory tracking controller for the Scripps and La Jolla maneuver}
    \label{fig:scripps_vel_cmd}
\end{figure}

\section{Summary and Conclusions}\label{sec:conclusion}

In this paper, a reduced order model (ROM) of a submarine was presented, along with adaptive autopilot controllers, path-following and trajectory-tracking algorithms. The hydrodynamic model is based on coefficients obtained from CFD simulations and its performance was demonstrated to compare favorably with CFD at lower speeds. Ongoing improvements are being made to the hydrodynamic model, particularly in the evaluation of secondary terms and the effect of boundaries.

Results from complex bathymetry maneuvers show that the path-following and trajectory tracking algorithms are able to safely achieve complex maneuvers. Further, the implementation of an adaptive controller in tandem with these algorithms can provide better maneuverability via adaptation to disturbances and non-ideal autopilot performance. The effect of the adaptive controller appears primarily to reduce the extrema of the errors, and in the case of the path-following and trajectory-tracking algorithms, reduce the time required to \textcolor{red}{converge to the path or trajectory}.  

In summary, it is shown that the path-following and trajectory-tracking algorithms are both very capable feedback control algorithms, and that each has a set of unique properties that suit to different scenarios. The path-following algorithm, being unconstrained by time, is more reflexive towards unknown or complex maneuvers, due to the fact that the velocity profile can be adjusted. The trajectory-tracking algorithm, on the other hand, is directly constrained by time, which means that the vehicle will achieve the desired maneuver at the exact prescribed final time, which is valuable in time-critical applications.

\section{Appendix}\label{sec:appendix}
\subsection{Path-Following Lyapunov Proof}\label{sec:appendix_pf}
Consider the Lyapunov candidate function \\
\centerline{$
    V_{PF} = \frac{1}{2}\bm{p_T}^T\bm{p_T}
$}\\
where $\bm{p_T}$ is defined in \eqref{eq:poserror}. \\
The derivative of the Lyapunov function is \\
$\begin{aligned}
\centerline{
$
    \dot{V}_{PF}=\bm{p_T}^T\bm{\dot{p}_T} $} \\ = [x_T,y_T]\left(-\bm{\omega_T}\times \bm{p_T}+\bm{R_W^T} 
    \begin{bmatrix}
    v \\
    0 
    \end{bmatrix}
    -
    \begin{bmatrix}
    ||\bm{p_d}'(\gamma)||\dot{\gamma} \\
    0 
    \end{bmatrix}
    \right)
\end{aligned}$ \\
where $\bm{\dot{p}_T}$ is defined in \eqref{eq:positionerrordynamics}. \\
Substituting \eqref{eq:gammad} for $\dot{\gamma}$, and noticing that $\bm{p_T}^T(-\bm{\omega_T}\times \bm{p_T}) = 0$ gives
\\
\centerline{
$
    \dot{V}_{PF}=[x_T,y_T]\left(\bm{R_W^T}
    \begin{bmatrix}
    v \\
    0 
    \end{bmatrix}
    -
    \begin{bmatrix}
    (v\bm{\hat{w}_1}+k_\gamma(\bm{p}-\bm{p_d}))^T\bm{\hat{t}_1} \\
    0 
    \end{bmatrix}
    \right)
$} 
In the case of ideal autopilot performance, i.e., $R_{W}^I=R_c$, we have 
\begin{equation} \label{eq:rwt}
\bm{R_W^T} = \begin{bmatrix}
    \frac{d}{(d^2+y_T^2)^1/2} & \frac{y_T}{(d^2+y_T^2)^1/2} \\
    \frac{-y_T}{(d^2+y_T^2)^1/2} & \frac{d}{(d^2+y_T^2)^1/2} 
    \end{bmatrix}
\end{equation} 
see Equation \eqref{eq:pfrot}. This leads to \\
\centerline{
$
    \dot{V}_{PF} = [x_T,y_T]\left(
    \begin{bmatrix}
    \bm{\hat{t}_1}^T\bm{\hat{w}_1} v \\
    \bm{\hat{t}_2}^T\bm{\hat{w}_1} v 
    \end{bmatrix}
    -
    \begin{bmatrix}
    \bm{\hat{w}_1}^T\bm{\hat{t}_1} v +k_\gamma(\bm{p}-\bm{p_d})^T\bm{\hat{t}_1} \\
    0 
    \end{bmatrix}
    \right)
$}
Recalling \eqref{eq:poserror}, we have $(\bm{p}-\bm{p_d})^T\bm{\hat{t}}_1 = x_T$, and noticing that $\bm{\hat{t}}^T_1\bm{\hat{w}}_1 = \bm{\hat{w}}^T_1\bm{\hat{t}}_1$, gives \\
\centerline{
$
    \dot{V}_{PF} = [x_T,y_T]
    \begin{bmatrix}
    -k_\gamma x_T \\
    \bm{\hat{t}_2}^T\bm{\hat{w}_1}v 
    \end{bmatrix}
$}
Further simplification results in \\
\centerline{$
    \dot{V}_{PF} =-k_\gamma x_T^2 + \bm{\hat{t}}_2^T\bm{\hat{w}}_1vy_T
$}
Noticing that $\bm{\hat{t}}_2^T\bm{\hat{w}}_1$ is a component of \eqref{eq:rwt}, the problem reaches its final simplification \\
\centerline{$
    \dot{V}_{PF} =-k_\gamma x_T^2 - \frac{vy_T^2}{||d^2+y_T^2||}
$}
Where $\dot{V}_{PF}(t)<0, \forall k_\gamma > 0, \forall t \in\mathbb{R}^2$

\subsection{Trajectory Tracking Lyapunov Proof}\label{sec:appendix_tt}
Consider the Lyapunov candidate function \\
\centerline{$
    V_{TT} = \frac{1}{2}\bm{e}_p^T\bm{e}_p$} \\
where $\bm{e}_p$ is defined in \eqref{eq:ttpositionerror} \\
The derivative of the Lyapunov function is \\
\centerline{$
    \dot{V}_{TT} = \bm{e}_p^T\bm{\dot{e}}_p
$} 
where $\bm{\dot{e}}_p$ is defined in \eqref{eq:ttpositionerrordyn} \\
\centerline{$
    \dot{V}_{TT} = \bm{e}_p^T\left(\bm{R}_T^I
    \begin{bmatrix}
    v_d \\
    0 
    \end{bmatrix} - 
    \bm{R}_W^I
    \begin{bmatrix}
    v \\
    0 
    \end{bmatrix}
    \right)
$} 
Recalling that $\bm{R_T^I}=\textcolor{red}{[\bm{\hat{t}_1},\bm{\hat{t}_2}]}$ and $\bm{R_W^I}=\textcolor{red}{[\bm{\hat{w}_1},\bm{\hat{w}_2}]}$ \\
\centerline{$
    \dot{V}_{TT} = \bm{e}_p^T(v_d\textcolor{red}{\bm{\hat{t}}_1}-v\textcolor{red}{\bm{\hat{w}}_1})
$} 
Combining Equations \eqref{asm:APTT} and \eqref{eq:ttrot} we have \\
\centerline{$
    \textcolor{red}{\bm{\hat{w}}_1} = \frac{k_p\bm{e}_p+v_d\textcolor{red}{\bm{\hat{t}}_1}}{||k_p\bm{e}_p+v_d\textcolor{red}{\bm{\hat{t}}_1}||}
$} 
and substituting \eqref{eq:ttvelocity} for $v$, $v\textcolor{red}{\bm{\hat{w}}_1}$ simplifies to \\
\centerline{$
    v\textcolor{red}{\bm{\hat{w}}_1} = (k_p\bm{e}_p+v_d\textcolor{red}{\bm{\hat{t}_1}})^T\textcolor{red}{\frac{k_p\bm{e}_p+v_d\textcolor{red}{\bm{\hat{t}}_1}}{||k_p\bm{e}_p+v_d\textcolor{red}{\bm{\hat{t}}_1}||}}
    \frac{k_p\bm{e}_p+v_d\textcolor{red}{\bm{\hat{t}}_1}}{||k_p\bm{e}_p+v_d\textcolor{red}{\bm{\hat{t}}_1}||}
$} 
Which reduces to \\
\centerline{$
    \textcolor{red}{v\bm{\hat{w}_1}=k_p\bm{e}_p+v_d\bm{\hat{t}_1}}
$} 
Continuing the derivation of $\dot{V}_{TT}$ \\
\centerline{$
    \dot{V}_{TT} = \bm{e}_p^T(v_d\textcolor{red}{\bm{\hat{t}}_1}-(k_p\bm{e}_p+v_d\textcolor{red}{\bm{\hat{t}}_1}))
$} 
\centerline{$
    \dot{V}_{TT} = \bm{e}_p^T(-k_p\bm{e}_p)
$} 
\centerline{$
    \dot{V}_{TT} = -k_p||\bm{e}_p||^2
$} 
Where $\dot{V}_{TT}(t) < 0, \forall k_p>0, \forall t \in \mathbb{R}^2$

\bibliographystyle{cas-model2-names}
\bibliography{references}

\end{document}